# The Effectiveness of "Pencasts" in Physics Courses

Nandana Weliweriya
Eleanor C Sayre
Dean A Zollman
Department of Physics, Kansas State University, Manhattan, KS 66506

Pencasts are videos of problem solving with narration by the problem solver. Pedagogically, students can create pencasts to illustrate their own problem solving to the instructor or to their peers. Pencasts have implications for teaching at multiple levels from elementary grades through university courses. In this article, we describe the use of pencasts in a university level upper-division Electromagnetic Fields course usually taken by junior and senior physics majors. For each homework assignment, students created and submitted pencasts of ordinary problems several days before the problem set was due. We compare students' performance in the class (grades for pencast submission excluded) with the pencast submission rate. Students who submitted more pencasts tend to do better in the course. We conclude with some practical suggestions for implementing pencasts in other courses.

**Introduction**

Solving Homework problems is a key element in every physics course at the high school and undergraduate levels. When instructors grade students' solutions, they just see sketches, equations, mathematical steps, and sometimes a few comments or sentences.  Important information about the process and reasoning behind student problem solving is missing. In order to teach students how to solve problems better, we need to study their problem solving process, not just their completed solutions.

A pencast is a digital recording of someone solving a problem in real-time, complete with both the things they write and audio of their narration as they write it. Students can complete a pencast using a smart pen to record the solutions to a problem.  Someone could also use video on their smart phone to record problem solving on paper while narrating the solution. Thus, pencasts enable us to capture problem solving steps along with vocal explanations.  If students use their cell phones to record their work on paper, the pencasts also capture their gestures (not whole body but hand movements).

Pencasts have become popular in comparison to traditional chalk-and-talk lessons or handwritten solutions due to the fact that the explanation is synchronized with relevant diagrams and/or mathematical steps. More importantly, users can replay the recording as needed. Students can quickly jump to a desired section rather than listening to hours of audio used with voice recorders. Furthermore, these videos may be preferred to traditional print instructional materials by students who prefer audio and visual input [1, 2].

Pencasts are becoming an interactive instructional tool in education, but their effectiveness has not yet been studied thoroughly [1], and faculty may be confused about how to implement them in their classes.  In the following sections, we outline how we use pencasts in our small-enrollment physics class, present some evidence of their effectiveness, and discuss technological issues in their implementation.

## Why use pencasts?

The Pearson Student Mobile Device Survey [3] in June 2015 found that more than eight in 10 (86%) college students regularly use a smartphone. Moreover, almost nine in 10 (87%) college students use a tablet, laptop, notebook or Chromebook computer every week in order to do their school work.  It is possible to take advantage of the ubiquity of these devices in an efficient way for academic work.

One of the best ways to understand a concept is to explain it to someone else. Developing pencasts is a way for students to learn by teaching i.e. a type of student-centered learning. For example, a new instructor plans a lecture on magnetic fields in matter.  He does not feel confident enough to teach it, and thus spends a substantial amount of time developing the groundwork and organizing the lecture materials. Finally, in lecture he explains the concepts smoothly along with extended examples. In preparation for teaching -- and in the moment of teaching -- he develops a deeper understanding of the physics. Asking students to regularly cover a lecture is fraught with pedagogical problems, so we cannot replicate this model to improve student learning more broadly in the class.

Pencasts represent a more practical model.  In preparing for and producing their problem solving process, students take advantage of learning by teaching.  Studies [4-6] have shown that the students gather extra knowledge through experience while preparing the pencasts themselves. Pencasts may also facilitate more time on task because it takes some time to prepare the problem and solve the problem while capturing the pencast.  Nonis and Hudson [7] found that the quantity of time spent studying has a positive impact on students' performances when students are able to concentrate.  Importantly, the benefits are larger than merely time on task, because performing aspects of pencasts is more than just mindless study. Students become better problem solvers when they explain their problem solving strategies to others [8].

From an evaluation perspective, pencasts are richer in content than a written solution on paper. When we pick good problems, we consider problems that include sense making, setting up, and explaining physics, but not problems with a lot of algebra, because algebra is time-consuming in a way that is unproductive for student sense-making. Compared to merely grading written work, pencasts capture more evidence of the students' thinking process as it has both audio and video. When illustrations are synchronized with solution steps, students' understanding of the course material is revealed. We have seen students use gestures frequently in sense making, explaining to fill a lack of words and to develop and share reasoning about abstract concepts. Videos capturing these rich moments give us an advantage that we cannot ever get from a written solution on paper. Grading pencasts is slightly slower than grading written work based on answers alone: the instructor needs to watch the pencast in order to grade it.  However, because students explain their reasoning throughout the problem, pencasts are easier to grade for the problem solving process, especially when students' solutions are atypical: the instructor doesn't need to guess why or how students went off the rails.

## Course setting

In this study we focus on the Electromagnetism I students. Twenty students were enrolled during the fall 2015 semester.  Grading pencasts -- like grading all aspects of a course -- is easier in smaller classes, so this course represents a good test case for new pedagogy.  This method is amenable to any class of this size, such as small introductory or high school classrooms; however, our introductory classes are too large in which to use these methods, so our examples are drawn from an upper-division course.

Usually the Electromagnetism I course is taken by junior or senior physics undergraduates. The goal of the course is to develop the theoretical tools to understand and apply Maxwell's equations in a broad variety of classical situations. The class met for four 50-minute periods each week for 15 weeks in the semester. It covered the first seven chapters of Introduction to Electrodynamics (4th Edition) by David J. Griffiths [9]. During the class, students worked in small groups on tutorials, and had full-class discussions on specific problems, mathematical methods, or physics concepts, where they were highly encouraged to work in groups.

In this course, homework (40% of grade) had two components: written and tutorial. Written homework was due weekly and tutorial homework was due two class days after the corresponding tutorial. The idea behind the homework was to encourage students to try to solve more examples related to in-class tutorials.

For each homework assignment, students submitted their pencasts. Overall, the pencasts represent 10% of their final grade in the course. Each pencast was due two days before the corresponding written homework. In each submission, students solved one of the homework problems assigned by the instructor. Students were allowed to use any device they wanted: smart pens, smart phones, tablets, or any sort of video camera. Students were required to capture their process of solving the problem (Figure 1) and a vocal explanation of the process. In order to give instructions on how to create pencasts the instructor uploaded a few pencast examples using different technologies.

As a trial run, students were asked to submit a short recording about themselves that was also used to identify and fix technical difficulties with production or uploading. We strongly recommend this trial: many students had technical difficulties at first (see "Technology Issues" section), and the trial run permitted us to test the technology without the pressures of solving physics problems.

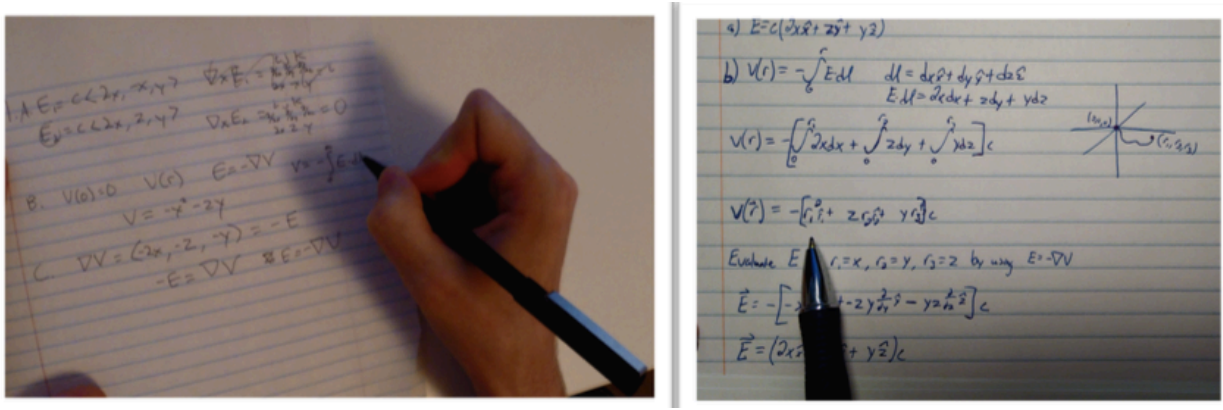

Figure 1: Screen shots of pencasts; two students solving problems

### Findings

The grading scale for this course had four elements: in-class work (small group problem solving, full class discussions and individual quizzes), which was 20% of their grade, written exams that were 30% of grade, homework (written and tutorial) was 40% of grade, and pencasts were 10% of their grade.

Figure 2 show that 60% of the students scored more than 80% in the class; they submitted all or nearly all of the pencasts. The students who submitted more pencasts tended to do better in the course, even after we calculated the portion of their grade resulting only from the other elements of the course.

We used a Wilcoxon rank sum test in looking at the final grade for all students as a function of the pencast submission rate. Figure 2 shows there is a low clump and high clump. Our data is non-parametric ordinal data that has two levels: low and high, therefore it is appropriate to use the Wilcoxon rank sum test. These clumps are different ($p < .0001$ by Wilcoxon rank sum test), and the high-submitters earned better grades than the low-submitters. We conclude that the students' submission of more pencasts correlates well with their overall performance in the course.

As an independent measure, the Colorado Upper Division Electrostatics Assessment (CUE) [10] was given to students at the start (Pre-test) and end (Post-test) of the course. When we analyzed student responses for pre-test and post-test, we found that in the pre-test most of the students tried to guess the answers or combine what they already knew to come up with something. The most common comments were "I don't know" or "I have no idea". However, in the post-test students mostly used equations and diagrams to represent the physical system and were much more successful at solving problems. Students frequently used a conceptual approach rather than long algorithmic calculations in their responses, a mark of conceptual understanding. The CUE is an overall assessment of their learning in the course, not one specifically tied to pencast production. Because pencast problems are drawn from their homework, we can't separate their learning on pencast topics versus non-pencast topics. Alas, we did not interview students about their impressions of doing pencasts.

We suggest that solving pencast problems as part of a larger effort to learn physics in the course helped students become more robust problem solvers.

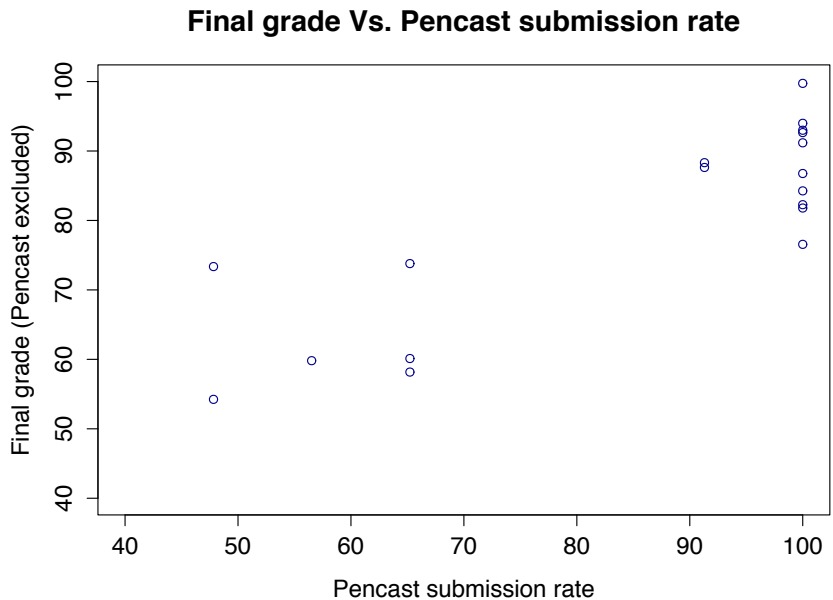

Figure 2: Final grade for all students vs. Pencast submission rate

## Technology issues and pencasts

The first pencast served as an experiment to identify the technical difficulties in producing videos and uploading them to the course management system for grading. Most of the students used their phones to record the videos and did not report any difficulties producing the videos, though some students were concerned that the audio quality might not be good enough (it was just fine -- consumer-grade audio is perfectly sufficient for this purpose). However, their phones recorded automatically in HD and the file size was too large to transfer directly from their phones to the course management system or via email. This issue was resolved by having the students export their videos to YouTube and then send the link to the instructor. YouTube videos can be set to one of three levels of privacy: completely open in which anyone with a link can view the video, and totally closed. Students need to choose the middle level in order to share links with the instructor.

For the small minority of students without smart phones, there were plenty of other technological solutions. Classroom tablets can record using their video cameras or special software to capture the screen as students draw. Students can use cheap video cameras (e.g. Web-cam) and download the video to a computer. Students can also use a Livescribe pen with special paper that is made specifically for this purpose. We offered to loan students the necessary equipment if they did not have their own although no students took us up on this offer. This is one important point we want to make: in order to make a pencast, the particular technology is not important, and sufficient technology is ubiquitously available.

## Implications for instructors

In order to learn, students need to actively engage in the learning process and participate in both classroom activities and homework. However, if students become stuck at a particular step within a problem, pencasts can be easily used to show students' mistakes or misconceptions, and become a starting point for correcting them.

Pencasts enable access to the instructor outside of class. If a struggling student needs further assistance, they can upload the recording of their work and receive feedback from the instructor. The instructor does not need to start an explanation from the beginning; he/she can address the exact point where the student needs help because pencasts capture evidence of the students' thinking process.

Students can also share their own pencasts with other students. This has far-reaching effects. In addition to the learning benefits of producing pencasts, reviewing each other's' pencasts may help struggling students understand the material better. Fostering collaboration in the classroom can also help students form a community of practice around solving physics problems, which may improve student retention in physics.

Furthermore, a video containing a student solving homework problems may help the instructor to understand how students sense and conceptualize the material. It allows the instructor to rescale according to the response of the students, and therefore may serve as an important formative assessment on the processes of problem solving.

## References:

1. J. Herold, T. Stahovich, H. Lin, and R.C. Calfee. "The effectiveness of "pencasts" as an instructional medium," paper presented at 2011 *ASEE Annual Conference & Exposition*, Vancouver, BC (June 2011). https://peer.asee.org/18518

2. D. M. Roesch, "Temporally contiguous pencast instruction promotes meaningful learning for dental and dental hygiene students in physiology," *J Dent Educ* 78, 51-55 (January 2014).
3. [http://www.pearsoned.com/wp-content/uploads/2015-Pearson-Student-Mobile-Device-Survey-College.pdf,](http://www.pearsoned.com/wp-content/uploads/2015-Pearson-Student-Mobile-Device-Survey-College.pdf) Pearson Student Mobile Device Survey 2015, National Report: College Students, (June 2015).
4. N. Esgi, "Comparisons of effects of students and teacher prepared screencasts on student achievement," *ESJ* 10(22), 1857-7881 (August 2014).
5. A. K. Shaffer and J. R. Schwebach, "Usefulness of Livescribe web recordings as supplemental resources for a large lecture undergraduate course," *J Coll Sci Teach,* 44(4) 54-60 (March/April 2015).
6. K. R. Green, T. Pinder-Grover, and J. M. Millunchick, "Impact of screencast technology: connecting the perception of usefulness and the reality of performance," *J Eng Educ* 101, 717-737 (October 2012).
7. S. A. Nonis and G.I. Hudson, "Performance of college students: impact of study time and study habits," *JEB* 85(4), 229-238 (2010).
8. K. Clemmer, J. McCallum, J. Phillips, and T. Zachariah, "Improving students' problem-solving competence with think-alouds," poster presented at the 2013 ISSOTL Conference (October 2013).
9. D. J. Griffiths, *Introduction to Electrodynamics*, 3rd ed. (Prentice Hall, Upper Saddle River, NJ, 2007).
10. [http://www.colorado.edu/sei/departments/physics_assessment.htm,](http://www.colorado.edu/sei/departments/physics_assessment.htm) Colorado Upper Division Electrostatics (CUE) Assessment.